\documentclass[conference]{IEEEtran}
\IEEEoverridecommandlockouts

\usepackage[ruled,linesnumbered]{algorithm2e}

\def\BibTeX{{\rm B\kern-.05em{\sc i\kern-.025em b}\kern-.08em
    T\kern-.1667em\lower.7ex\hbox{E}\kern-.125emX}}
    
\usepackage{CJKutf8}  

\usepackage{xcolor}
\usepackage{pgfplots}
\usepackage{tikz}
\usepackage{comment}

\usepackage{filecontents}

\definecolor{armygreen}{rgb}{0.29, 0.33, 0.13}

\usepackage{fancyhdr} 
\usepackage{hyperref}
\hypersetup{
     colorlinks = true,
     linkcolor = blue,
     anchorcolor = red,
     citecolor = magenta,
     filecolor = blue,
     urlcolor = black,
}

\usepackage{multirow}
\usepackage{cite}
\usepackage{amsmath,amssymb,amsfonts}

\usepackage{textcomp}
\usepackage{xcolor}

\usepackage{subfigure}
\usepackage{pgfplots}
\usepackage{filecontents}

\usepackage{graphicx,subfigure}

\definecolor{bblue}{HTML}{4F81BD}
\definecolor{rred}{HTML}{C0504D}
\definecolor{ggreen}{HTML}{9BBB59}
\definecolor{ppurple}{HTML}{9F4C7C}

\newtheorem{defi}{Definition}

\newcommand{\cev}[1]{\reflectbox{\ensuremath{\vec{\reflectbox{\ensuremath{#1}}}}}}

\usepackage{listings}

\usepackage{blindtext}
\usepackage{etoolbox,xstring,mfirstuc,textcase}
\usepackage{booktabs}
\usepackage{pgfplots}

\usepackage{tabularx,booktabs,makecell,multirow, caption}
\usepackage{enumitem} 

\newcolumntype{L}{>{\arraybackslash}X}

\usepackage{xcolor}
\usepackage{tikz}
\usepackage{pgfplots}

\definecolor{findOptimalPartition}{HTML}{D7191C}
\definecolor{storeClusterComponent}{HTML}{FDAE61}
\definecolor{dbscan}{HTML}{ABDDA4}
\definecolor{constructCluster}{HTML}{2B83BA}

\usepackage[utf8]{inputenc}
\usepackage{listings}
\begin{document}
\title{A Referable NFT Scheme}


\author{\IEEEauthorblockN{Qin Wang\IEEEauthorrefmark{1}, Guangsheng Yu\IEEEauthorrefmark{1}, Shange Fu\IEEEauthorrefmark{2}, Shiping Chen\IEEEauthorrefmark{1}, Jiangshan Yu\IEEEauthorrefmark{2}, Xiwei Xu\IEEEauthorrefmark{1}
}

\IEEEauthorrefmark{1}  \textit{CSIRO Data61, Australia}\\
\IEEEauthorrefmark{2} \textit{Monash University, Australia}
}

\IEEEoverridecommandlockouts

\IEEEpubid{\makebox[\columnwidth]{978-8-3503-1019-1/23/\$31.00~\copyright2023 IEEE \hfill} \hspace{\columnsep}\makebox[\columnwidth]{ }}

\maketitle


\begin{abstract}
Existing NFTs confront restrictions of \textit{one-time incentive} and \textit{product isolation}. Creators cannot obtain benefits once having sold their NFT products due to the lack of relationships across different NFTs, which results in controversial profit sharing. This paper proposes a referable NFT solution to extend the incentive sustainability of NFTs. We construct the referable NFT (rNFT) network to increase exposure and enhance the referring relationship of inclusive items. We introduce the DAG topology to generate directed edges between each pair of NFTs with corresponding weights and labels for advanced usage. We accordingly implement and propose the scheme under Ethereum Improvement Proposal (EIP) standards, indexed in \textcolor{armygreen}{EIP-5521}. Further, we provide the mathematical formation to analyze the utility for each rNFT participant. The discussion gives general guidance among multi-dimensional parameters. The solution, as a result, shape the recognition of potential values hidden in isolated NFTs and raise the interest of communities toward the discovery of NFT derivatives. To our knowledge, this is the first study to build a referable NFT network, explicitly showing the virtual connections among NFTs. 
\end{abstract}

\smallskip

\section{Introduction}
\label{sec-intro}

Non-fungible tokens (NFTs)~\cite{wang2021non} is built as EIP-721~\cite{erc721} to exchange unique digital assets in the Ethereum platform~\cite{wood2014ethereum}. It digitalizes on-chain assets into tokens, and specifies each token with a unique identifier $\textit{tokenID}$ within smart contracts. This significantly encourages and extends the on-chain token exchange from fungible to non-fungible ones which, not surprisingly, is now leading a wave of the next generation of a wide variety of applications covering virtual collectables, online tickets, digital arts, etc. Besides, NFT can be seamlessly incorporated with the protocols in decentralized finance (DeFi) \cite{werner2022sok} and the governance in blockchain communities \cite{kiayias2022sok}. To date\footnote{Data captured from \url{https://nonfungible.com/reports} [Q1-Q3 2022].}, a total of $33,651,380$ of NFT sales has led to the traded volume reaching up to $20$ billion USD. The phenomenal trading activities reflect a sharp shift from traditional markets into the new Web3 world \cite{wang2022exploring}.

However, the lack of relationships between an NFT and its owner in existing NFT standards could result in controversial profit sharing if the creation of a new NFT refers to previous NFTs. The trades of NFTs are \textit{one-time} in transferring and \textit{isolated} across different users. An NFT creator cannot continuously obtain profits for his intellectual property once sold. The permanent reference relationship between NFTs, thus, becomes increasingly important. The reference topology will assist in establishing a sustainable incentive mechanism to economically inspire more and more users to devote their contributions to using, creating, and promoting NFTs. To fill the gap, we propose \textcolor{armygreen}{EIP-5521} that defines a referable NFT token (rNFT) standard. It extends the static NFT into a virtually extensible NFT network. Users under clear ownership inheritance do not have to create work completely independent from others, avoiding reinventing the same wheel. Here, we summarise \textit{\textbf{contributions}} as follows:

\noindent\hangindent 1em \textit{\textbf{Protocol Design}} (Sec.\ref{sec_system}).  Our referable NFT scheme establishes the reference of ancestors when minting a new NFT with the reference relationship including both the \textit{referring} and \textit{referred} relationships, thus shaping a Direct Acyclic Graph (DAG) to represent the reference information. The scheme can provide reliable, trustworthy, and transparent historical records that every agreement can refer to in terms of profit sharing. Meanwhile, users are allowed to query, trace and analyze their relationship.

\noindent\hangindent 1em \textit{\textbf{Standard Implementation}} (Sec.\ref{sec_system}).  We implement a very succinct version and propose it to the Ethereum standard Git repository, indexed by \textcolor{armygreen}{EIP-5521}. Our simplified proposal is a smart-contract driven standard used as a system-level function on blockchains. It builds and retains the backward reference relationship between old works and new works and is compatible with existing mainstream standards. At the time of writing, rNFT has attracted consistent discussions and attention on forums \cite{forum} and by several in-production projects (e.g., Briq \cite{briq}).

\noindent\hangindent 1em  \textit{\textbf{Incentive Analysis}} (Sec.\ref{sec_incentive}). We provide a clear and exact mathematical expression of the utility function of an rNFT user. rNFT helps to integrate multiple upper-layer incentive models and also involves multi-dimensional parameters. We accordingly analyze the payoff function under different parameters. Chasing an optimal strategy for players is possible but without certainty. 

\noindent\hangindent 1em  \textit{\textbf{Further Discussions}} (Sec.\ref{sec_discussion}). We provide an in-depth discussion of our proposed scheme in terms of its opportunities and challenges. rNFT can promote a wide range of new NFT derivatives that requires historical relations but also confronts many pending aspects for discovery such as multi-contract or CC0 license development.

\smallskip
\noindent\textbf{An Intuitive Instance.} The constructed relationship topology between each NFT forms a DAG. By adding the \textit{referring} indicator, users can mint new NFTs (e.g. C, D, E) by referring to existing NFTs (e.g. A, B), while \textit{referred} enables the referred NFTs (A, B) to be aware that who has quoted it (e.g. $A\gets D$; $C\gets E$; $B\gets E \& A\gets E$). \textit{createdTimestamp} is an indicator, based on block timestamp, used to show the creation time of NFTs (A, B, C, D, E).

\smallskip
\noindent\textbf{Extending NFT Derivatives.} Similar to traditional financial derivatives that are based on stocks, commodities, or other underlying assets, NFT derivatives mainly refer to financial instruments that center around the value of an underlying NFT asset. The key feature of these derivatives is to allow users to invest in NFTs without actually holding the physical NFT asset in the forms such as options, futures, swaps, and other financial instruments that are used for the trading or hedging of the underlying NFT asset's price. However, this requires a relatively comprehensive history of records and credits. rNFT provides a foundation for atop derivative protocols, as siting to offer more complex investment strategies and risk management tools by establishing a reliable relationship among multiple linkable NFTs. The extension of NFT derivatives is an ongoing process and our solution gives an example for better connections.

\smallskip
\noindent\textbf{Integration.} Our scheme can be further integrated with mainstream techniques. We list two examples as the enlightenment. Firstly, the new rNFT can be combined with \textit{Graph Neural Network (GNN)}~\cite{scarselli2008graph} to achieve efficient queries and predictions of NFT. This is particularly useful for conducting automated labeling by NFT platforms such as OpenSea \cite{opensea}, or for NFT viewers to find more NFTs of the same kind. Secondly, AI detection \cite{kumar2010use} can be imported into the system for plagiarism detection, path recommendation, and strategy optimization. Due to the blur edges between the original artwork and copy-work, figure classification and identification in artificial intelligence is a fundamental tool for the detection of a rapidly growing NFT markets. 

\smallskip
\noindent\textbf{Backward Compatibility.} As in Sec.\ref{sec_system}, an rNFT implements the existing ERC-721 interfaces~\cite{erc721} to enable the backwards compatibility. This allows the proposed rNFT to be seamlessly implemented on existing blockchain platforms such as Ethereum~\cite{wood2014ethereum} where those existing ERC-721-compatible NFTs, such as the composable NFT: EIP-998~\cite{erc998}, can be referred by any subsequent rNFTs.

\smallskip
\noindent\textbf{Related Standards (Tab.\ref{tab-tokenstandard}).} A series of NFT-related standards\footnote{Captured from Ethereum ERCs \url{https://eips.ethereum.org/erc} [Dec 2022].} haves been intensively proposed. We highlight the ones entering the \textit{final} and \textit{last call} stages due to their deterministic probability of being accepted by communities. (EIP-)3525 proposes the concept of semi-fungible token by introducing a triple scalar $\langle id, slot, value \rangle$. ID acts the way as the 721 standards while additionally adding a quantitative feature \textit{value}. 1155 proposes a multi-token standard that can combine either fungible tokens, non-fungible tokens, or other configurations (e.g. semi-fungible tokens). 2981 focuses on the royalty payment transferred between a buyer and a seller voluntarily. 4907 and 5006 propose rental NFTs by extending original settings with an additional role of \textit{user} and a timing feature of \textit{expire}. A user can only use an NFT within the \textit{expire} duration, rather than transferring it. 2309 implements a consecutive transfer extension that enables batch creation, transfer, and burn methods by contract creators. The users can quickly and cheaply mint at most $2^{256}$ tokens within one transaction. 4400 extends 721's customer role of being able to, for instance, act as an operator or contributor. 5007 introduces the functions of \textit{startTime} and \textit{endTime} set a valid duration that automatically enables and disables on-chain NFTs. 4906 extends the scope of \textit{Metadata} that allows tracking changes by third-party platforms. 5192 proposes the idea of \textit{soulbound} that binds an NFT to a single account, achieving special applications such as non-transferable and socially-priced tokens. Besides, we also provide more related standards with indirect impacts on NFTs in Tab.\ref{tab-tokenstandard}).

\begin{table}[!hbtp]
 \caption{Summary of Token Standards} 
 \label{tab-tokenstandard}
  \centering
 \resizebox{1\linewidth}{!}{
 \begin{tabular}{lclr}
    \toprule
    \multicolumn{1}{c}{\textbf{Standard}}  
     & \multicolumn{1}{c}{\textbf{Platform}}  & \multicolumn{1}{c}{\textbf{Feature}} &
    \multicolumn{1}{c}{\textbf{Application}} \\
    \midrule
   EIP721  & Ethereum  & Non-Fungible Token & Artwork/IP   \\
   EIP777  & Ethereum  & Token Approval &  \\
   EIP1155 & Ethereum  & Adding an attribute for groups & Game   \\
   EIP3525 &  Ethereum  & Additional attribute for semi-fungible & Finical Market  \\
   EIP2981 &  Ethereum  & Retrieving the royalty payment info & Royalty Payments    \\
   EIP4907 &  Ethereum  & Adding a new role and timer &  Rental Market \\
   EIP5006 &  Ethereum  &  Adding the new role of user & Rental Market   \\
   EIP2309 &  Ethereum  & Consecutive token identifiers &   Consecutive event \\
   EIP4400 &  Ethereum  & Extend the consumer's functionalities & Authorization   \\
   EIP5007 &  Ethereum  & On-chain time management & Lend Market   \\
   EIP4906 &  Ethereum  & Enable token metadata's update  & Upgrade   \\
   EIP5192 &  Ethereum  & Bound to a single account & Soulbound Items   \\
   \textit{\textbf{This work}} &  \textbf{Ethereum}  & \textbf{Referable connections} &  \textbf{NFT Graph} \\
   \midrule
   EIP20   & Ethereum & Token API / Fungible Token & Vote/ICO  \\
   EIP223 & Ethereum  & Token Recovery  &  \\
   EIP998  & Ethereum  & Composable Non-Fungible Token &  Game/Ownership   \\
   EIP1238 &  Ethereum  & Non-Transferrable Non-Fungible Token & Badge  \\
   EIP1594 &  Ethereum  & Security Token Standard & Financial Securities  \\
   EIP1400 &  Ethereum  & Security Token Standard &  Securities    \\
   EIP1404 & Ethereum & Simple Restricted Token Standard & Securities  \\
   EIP1410 &  Ethereum & Partially Fungible Token Standard   \\
   EIP1462 &  Ethereum  & Base Security Token &  Securities  \\
   \midrule
   BEP20 &  Binance  & Fungible Token  & Vote/Wrap Token   \\
   BEP721 &  Binance  & Non-Fungible Token & IP Products  \\
   ARC721 &  Avalanche  & Fungible Token & Wrap Other Tokens   \\
   
   \bottomrule
  \end{tabular}
 }
\end{table}

\section{Protocol Design and Implementation}
\label{sec_system}

In this section, we present the rNFT protocol, construction, and implementation, respectively.  

\begin{figure*}[h]
    \centering
    \includegraphics[width=0.9\textwidth]{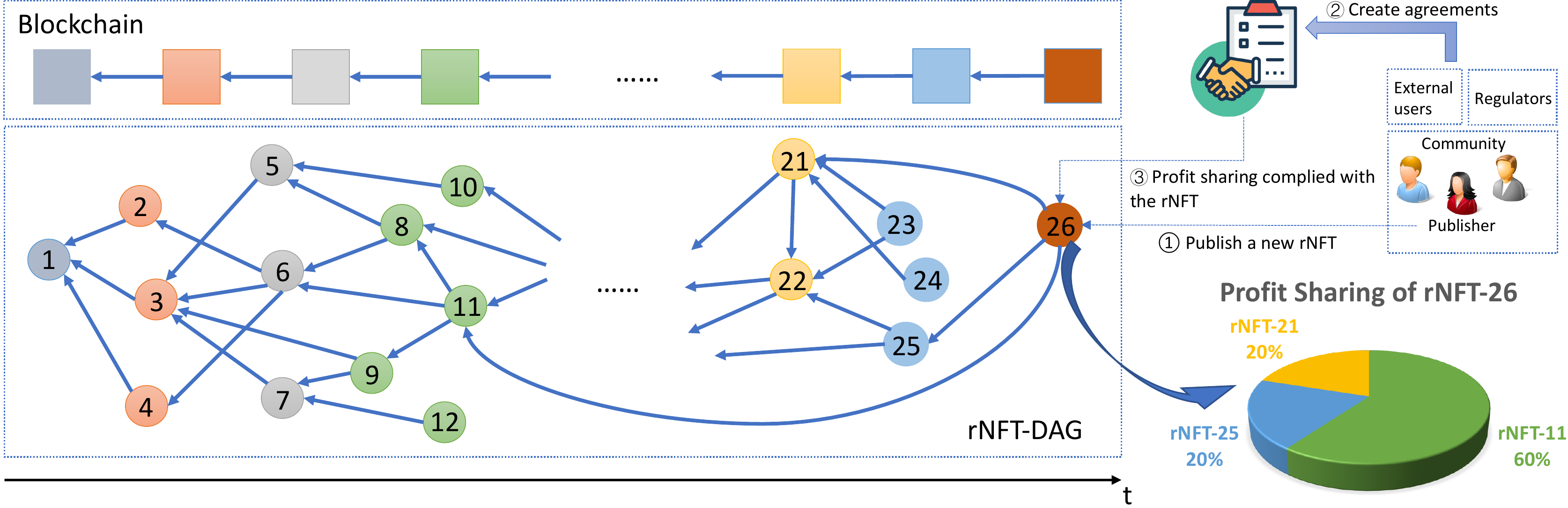}
    \caption{This figure shows the system overview throughout the item authentication workflow based on the proposed rNFT.}
    \label{fig: system-arch}
\end{figure*}

\smallskip
\noindent\textbf{Notations.} The notations used in this paper are listed as follows. $t$ is the timestamp of the block header. $\mathcal{T}_{\mathcal{R}}$ means a transaction that packs a particular rNFT and $\mathcal{B}_{h, t}$ is a block at the height of $h$ released at time $t$.
$\mathcal{R}$ represents the set of rNFTs, while $\mathcal{R}_{i, t, \Theta}$ is an rNFT released by user-$i$ at time $t$ with a set of $\Theta$ that contains both the referring and referred relationship, respectively, i.e., $\vec{\Theta}$ and $\cev{\Theta}$. Specifically, $\vec{\Theta}$ represents a collection of referring relationships made by an rNFT $\mathcal{R}_{i, t, \Theta}$ that connects several previous rNFTs such as $\mathcal{R}_{i, t', \Theta}$ where $t'<t$, denoted by the blue arrows in the rNFT-DAG, and
$\cev{\Theta}$ follows the same notation with $t'>t$, and is accordingly updated when this gets referred by any future rNFTs.
Note that the rNFT with $|\Theta|=0$ generally indicates an (claimed to be) original item. $\mathbb{A}_{i, j, k, ...}$ is an agreement involving user $i$, $j$, $k$, etc. The set of rNFTs $\mathcal{R}_{i, t, \Theta}$ constitutes a growing rNFT-DAG $\mathbb{D}$ as depicted at the bottom of Fig.~\ref{fig: system-arch}. Here, $\mathbb{D}$ merely represents a logical relationship graph (or topology) rather than an actual DAG network. Matched color between the nodes $\mathcal{R}$ in rNFT-DAG $\mathbb{D}$ and the blocks $\mathcal{B}$ on-chain indicates each $\mathcal{R}$ is expected to be packed in the corresponding block with the same color as time goes by.

\smallskip
\noindent\textbf{Protocol Design.} The rNFT protocol aligns with the narrative of any same kind of smart contract (SC-)supported protocols. We extract the core processes and highlight our contributions. 

\noindent\hangindent 1em \textbf{\textit{Parameter Setup:}}  The algorithm is used to create initial identities for users. The algorithm takes as input the security parameter $\kappa$, and outputs a key pair $(sk, pk)$ and corresponding \textit{address}es $addr$. 

\begin{equation*}
\centering
\begin{aligned}
sk \gets \mathsf{KeyGen_{sk}}(1^{\kappa}), pk \gets \mathsf{KeyGen_{pk}}(sk), \\ addr \gets \mathsf{AddressGen}(pk).
\end{aligned}
\end{equation*}

\noindent\hangindent 1em \textbf{\textit{Transaction Creation:}} This algorithm takes as input the user's private key $sk$, the transaction metadata $md$, the payload $pd$, and outputs the created transaction $\mathcal{T}$ that contains the signature $sig$ from the creator.

\begin{equation*}
\begin{aligned}
sig \gets \mathsf{Sign}(sk, md, pd), \\ \mathcal{T} \gets \mathsf{TranGen}(sig, md, pd).
\end{aligned}
\end{equation*}

\noindent\hangindent 1em \textbf{\textit{Contract Execution:}} The algorithm takes as input the created transaction $\mathcal{T}$, and the state $s$, and $contract$ that contains the operating logic, and outputs the transited state $s$.
\begin{equation*}
s \gets \mathsf{ContractExec}(s, \mathcal{T}, contract),
\end{equation*}
The state $s$ includes the newly minted rNFT $\mathcal{R}_{i, t, \Theta}$ with the current state of referred relationships $\cev{\Theta}$. Here, a pre-released rNFT $\mathcal{R}_{i, t, \Theta}$ is expected to be included in a transaction $\mathcal{T}_{\mathcal{R}}$ that will be subsequently packed by a valid block $\mathcal{B}_{h,t}$ on a growing chain with finalization at the height of $h$ at time $t$. 

\noindent\hangindent 1em \textbf{\textit{State Consensus:}} This algorithm takes as input the transaction $\mathcal{T}$, the smart contract $contract$ and the state $s$ to be transited, and outputs the confirmed state $s^{\prime}$, and the confirmed transaction $\mathcal{T}^{\prime}$.

\begin{equation*}
(s^{\prime}, \mathcal{T}^{\prime}) \gets \mathsf{Consensus}(s, \mathcal{T}, contract).
\end{equation*}

We can observe that as soon as one from the community as the publisher $i$ at time $t$ publishes a new rNFT $\mathcal{R}_{i, t, \Theta}$ with a bunch of referring relationships collected in $\Theta$, and the corresponding emitted transaction $\mathcal{T}_{\mathcal{R}}$ is finalized by a valid block $\mathcal{B}$ on the blockchain, any agreement $\mathbb{A}$ will be able to enjoy a reliable, trustworthy, and transparent historical record about the items, referring relationships, and profit sharing to guarantee the fair trading later on.

\smallskip
\noindent\textbf{Implementation.} We implement the rNFT scheme by following the ERC standards, and propose the entire solution as indexed in \textcolor{armygreen}{EIP-5521}, which is an extension of the ERC-721 protocol. Our solution adds two types of referable indicators, \textit{referring} and \textit{referred}, and a time indicator \textit{createdTimestamp} in a new structure \textcolor{teal}{$\mathsf{Relationship}$}. The \textit{referring} indicator (implemented by \textcolor{teal}{$\mathsf{setNodeReferring}$} and \textcolor{teal}{$\mathsf{referringOf}$}) enables a user, as the child, to inherit the ancestor NFTs. The \textit{referred} indicator (by \textcolor{teal}{$\mathsf{setNodeReferred}$} and \textcolor{teal}{$\mathsf{referredOf}$}) makes a parent NFT builder be able to check who has referenced him. Also, we provide two optional advance indicators \textit{labels} and \textit{profitSharing} for additional functions like incentive distribution and attributes establishment. The time indicator is a block-level timestamp that assists in solving disputes. Specifically, we set five \textbf{\textit{indicators}} as the parameter to implement the proposed scheme, as shown in Algorithm.\ref{algorithm1}

\begin{algorithm} 
\caption{\textbf{rNFT Standard Interfaces}}\label{algorithm1}
\BlankLine
 \textbf{interface} ERC721 \{ \\
 \quad    function \textcolor{teal}{$\mathsf{ownerOf}$}(uint256$\_$tokenI) external view returns (address); \\
 \quad    function \textcolor{teal}{$\mathsf{transferFrom}$}(address$\_$from, address$\_$to, uint256$\_$tokenId) external payable;  ... \} \\
      
 \textbf{interface} \textcolor{armygreen}{EIP-5521} \{ \\
 \quad    function \textcolor{teal}{$\mathsf{setNode}$}(uint256 tokenId, uint256[] memory $\_$tokenIds) external; \\
 \quad    function \textcolor{teal}{$\mathsf{referringOf}$}(uint256 tokenId) external view returns(uint256[] memory);\\
 \quad    function \textcolor{teal}{$\mathsf{referredOf}$}(uint256 tokenId) external view returns(uint256[] memory); ... \} \\
      
 \textbf{struct} Relationship \{ \\
 \quad         uint256[]  \textcolor{armygreen}{referring};\\
 \quad         uint256[]  \textcolor{armygreen}{referred};\\
 \quad         uint256  \textcolor{armygreen}{createdTimestamp}; \\ 
 \quad    \textcolor{armygreen}{extensible parameters} \quad  \} \\

 \textbf{function}  \textcolor{teal}{$\mathsf{setNode}$}(uint256 tokenId, uint256[] memory $\_$tokenIds) public virtual override \{ \\
  \quad      \textcolor{teal}{$\mathsf{setNodeReferring}$}(tokenId, $\_$tokenIds);\\
  \quad      \textcolor{teal}{$\mathsf{setNodeReferred}$}(tokenId, $\_$tokenIds);\quad \} \\
    
\label{algo:rnft}
\end{algorithm}

\begin{itemize}
    \item[-] \textit{referring}: an out-degree indicator,  showing the users this NFT refers to.
    \item[-] \textit{referred}: an in-degree indicator, showing the users who referred this NFT. 
    \item[-] \textit{createdTimestamp}: a time-based indicator, comparing the timestamp of mint.
    \item[-] \textit{labels} (Advanced): a list recording the attributes or categories of the rNFT.
    \item[-] \textit{profitSharing} (Advanced): a list recording the profit sharing of \textit{referring}, thus the size of \textit{profitSharing} equaling to that of \textit{referring}.
\end{itemize}

Here, we can observe that customizable attributes are also added following the referable indicators, for instance, a string array of \textit{labels} can store a series of attributes given to a particular NFT, such as \textit{artwork}, \textit{song}, \textit{movie}, \textit{animation}, \textit{subtitle}, \textit{screenshot}, etc. Another optional but useful one could be a list recording the profit sharing among the owners whose NFTs have referred to this NFT. By complying with the record of profit sharing, fair and transparent on-chain or off-chain agreements can be achieved. Then, we also have six corresponding \textbf{\textit{methods}} for the implementation.

\begin{itemize}
    \item[-] \textcolor{teal}{$\mathsf{safeMint}$}: mint a new rNFT.
    \item[-] \textcolor{teal}{$\mathsf{setNode}$}: set the referring list of an rNFT and update the referred list of each one in the referring list.
    \item[-] \textcolor{teal}{$\mathsf{setNodeReferring}$}: set the referring list of an rNFT.
    \item[-] \textcolor{teal}{$\mathsf{setNodeReferred}$}: set the referred list of the given rNFTs.
    \item[-] \textcolor{teal}{$\mathsf{referringOf}$}:  get the referring list of an rNFT.
    \item[-] \textcolor{teal}{$\mathsf{referredOf}$}: get the referred list of an rNFT.
\end{itemize}

Based on the above functions, \textcolor{armygreen}{EIP-5521} makes involved rNFTs form a DAG that can explicitly show their relationship and entanglement. The protocol increases the incentive of existing NFTs where an NFT owner can obtain more revenues from the followers. In the next part, we give our theoretical analysis of incentive optimization within the rNFT ecosystem.


\section{Incentive Analysis}
\label{sec_incentive}

In this section, we discuss the incentive analysis. In total, we majorly consider three orthogonal \textit{principles}. From the \textit{horizontal} view (count by \textit{in/out degree}), an rNFT that have been referred more times should gain much more incentives. From the \textit{vertical} view (count by \textit{depth}), an rNFT with a deeper reference index should accumulate more revenues. For the \textit{profit distribution}, we additionally introduce the \textit{weight} to ensure the distribution will be ended within finite rounds. We also set $\sigma$ to measure the descending rate according to its depth $d$ for the subsequent users. Specifically, we set the following parameters:

\begin{itemize}
    \item[-] \textit{Referring bandwidth ($\mathcal{I}$)}: measure the number of rNFTs that refer to this NFT at the same block height.
    \item[-] \textit{Referring depth ($d$)}: measure the number of attached children in every single path of the branch.
   Here, $d$ is calculated by $d=\Delta h = h_\text{expire}-h_\text{init}$.
    \item[-] \textit{Threshold ($\lambda$)}: set the initial pay rate (anchoring its initial value $p_0$) when minting a new rNFT.
    \item[-] \textit{Weight ($\vec{w}$)}: set the weight for each referring edge (w.r.t. referring list $\vec{\Theta}$).
    \item[-] \textit{Rate ($\sigma$)}: adjust the descending rate of the income as the depth increases. 
\end{itemize}

We consider the payoff function of each single existing rNFT user.

\begin{defi} The payoff function $\mathbb{U}$ of a user $i$ releasing an rNFT $\mathcal{R}_{i,h,\Theta}$ at the block height $h$ is made by
$$\mathbb{U}({\mathcal{R}_{i,h,\Theta}})= \mathcal{I}_{\mathcal{R}_{i,h,\Theta}} - \mathcal{O}_{\mathcal{R}_{i,h,\Theta}},$$
where $\mathcal{I}$ represents accumulated income revenues during the following rounds while $\mathcal{O}$ is the cost of minting this rNFT.
\end{defi}

For the outcome, $\mathcal{O}_{i,h,\Theta}$ indicates the costs paid to the network. We can see that $\mathcal{R}_{i,h,\Theta}$ is minted at the round $h$ (according to its block height), with the minting fees of $p_0$. Here, he can optionally pay a part of (measured by $\lambda$) the entire charge at $\lambda p_0$. But accordingly, the rest of the payment needs to be completed within the next $d$ rounds, where $d$ is calculated by $d=g(1-\lambda)$ and $g$ is a constant. The delayed payment will accompany by an interest rate $r$. Instead of relying on a simple average assignment, we relate $r$ with individual weights $w_i$ of each rNFT. Specifically, we define $w_0$ as the weight of \textit{self-references} with $w_i$ being the weights of all other \textit{cross-references} $\forall i>0$. 

We first give details for calculating the initial price $p_0$. The initial paid $p_0$ is \textit{one-off}. It contains a ratio list, where $d$ previous rNFTs are referred with $w_i$ being the ratio of profit sharing for $i$-th rNFT in the referring list $\vec{\Theta}$. Note that having to allocate the $w_0$ implies that a constant expense $\hat{\mathcal{O}}>0$ is mandatory for releasing each rNFT regardless of the size of the referring list $\vec{\Theta}$. We obtain the calculation of $p_0$ as

\begin{equation*}
\begin{aligned}
p_0
= & \hat{\mathcal{O}}\times\lbrack w_0, w_1, w_2, \dots, w_{|\vec{\Theta}|}\rbrack, \text{ with}\sum_{i=0}^{|\vec{\Theta}|} w_i=1 \text{ and }w_i \in \mathbb{R}_0^+.
\end{aligned} 
\end{equation*}

Accordingly, the outcome function with compound interest is stated as
\begin{equation*}
\begin{aligned}
\mathcal{O}_{\mathcal{R}_{i,h,\Theta}} 
= & \lambda p_0 + (1+r) \frac{p_0 (1-\lambda)}{d} + (1+r)^2 \frac{p_0 (1-\lambda)}{d}   + ...  \\
  & +  (1+r)^d \frac{p_0 (1-\lambda)}{d} \\ 
= & \lambda p_0 + \sum_{j=h}^{h+d} (1+r)^d  \frac{p_0 (1-\lambda)}{d}, \\
& \text{ where } r \varpropto \alpha{\sum_{i=1}^{|\vec{\Theta}|} w_i} \text{ with }\alpha \in [0,1].  \\
\end{aligned} 
\end{equation*}

For the income, $\mathcal{I}_{i,h,\Theta}$ can obtain the accumulated rewards from all participants' payments. The income is to collect same-height income (the \textit{horizontal} dimension) iteratively for all valid rounds (the \textit{vertical} dimension), where each round of income relates to actual participants $\mathcal{I}$. Also, we use the descending rate $\sigma$ to adjust the valid rounds for profit distribution, which is set to be an inversely proportional relation. Based on the targeted rNFT's position on-chain, we calculate its accumulated incomes from each following round. At the round $h+1$, the relevant new set of minted rNFT that refer to $\mathcal{R}_{i,h,\Theta}$ is $\mathcal{I}_{(i+1)}$ and the descending rate for each round is given by $\sigma=\frac{1}{w_0+\beta} \text{ with }\sigma \in [0,1] \text{ and $\beta$ is a non-zero constant}$.

We note that the settings of $r$ and $\sigma$ can be effectively used to adjust the trade-off between the income $\mathcal{I}_{i,h,\Theta}$ and outcome $\mathcal{O}_{i,h,\Theta}$, hence leading to the payoff function $\mathbb{U}$. The higher weight of the self-reference, the lower the interest rate $r$ can be offered to encourage creative and original works. The higher weight of the cross-reference, the lower descending rate $\sigma$ can be offered to works with various cross-references which are commonly acknowledged as mature products.


Then, we calculate round revenues. The revenue of the round $h+1$ is calculated as $k\sigma|\mathcal{I}_{(i+1)}|$ where $k$ is a constant. Similarly, the income at the round $h+2$ is $k\sigma^2|\mathcal{I}_{(i+2)}|$. We thereby have the income function as

\begin{equation*}
\begin{aligned}
\mathcal{I}_{\mathcal{R}_{i,h,\Theta}} 
= & k\sigma |\mathcal{I}_{(h+1)}| + k\sigma^{2} |\mathcal{I}_{(h+2)}| + ... + k\sigma^{d} |\mathcal{I}_{(h+d)}|  \\
=& \sum_{j=h}^{h+d} k \sigma^{d} |\mathcal{I}_{(h+d)}|.\\
\end{aligned} 
\end{equation*}
\textit{where $d$ is the depth for measuring valid rounds of profit distribution, calculated by $d=g(1-\lambda)$.} Based on the above equations, we obtain the payoff function as

\begin{equation*}
\begin{aligned}
\mathbb{U}_{\mathcal{R}_{i,h,\Theta}} 
= & \sum_{i=h}^{h+d} k\sigma^{d}|\mathcal{I}_{(h+d)}| -  \lambda p_0 - \sum_{i=h}^{h+d}(1+r)^d \frac{p_0 (1-\lambda)}{d}  \\
= & \sum_{i=h}^{h+d} k\sigma^{d}|\mathcal{I}_{(h+d)}| -  p_0\left[ \lambda+  \sum_{i=h}^{h+d}(1+r)^d g^{-1} \right]  \\
& \propto  \, \,  d \lbrack \sigma^d |\mathcal{I}| - (1+r)^{d}\rbrack.
\end{aligned} 
\end{equation*}
Intuitively, we can conclude that the payoff revenues of each minted rNFT are positively proportional to the accumulative participants (namely, $\sum_d|\mathcal{I}|$) or valid references ($|\Theta|$), whereas negatively proportional to the round interest $r$ if not set to be $0$. The more a user paid for the initial price (say $\lambda$), the less round he needs to pay to the previous rNFTs. Similarly, the higher descending rate (cf. $\sigma$) is set, the fewer rounds ($d$) can last for creating benefits. However, the overall payoff function relates to all parameters including $|\mathcal{I}|$, $\lambda$, $d$, $w$ and $\sigma$ is complicated. It is unpredictable to decide the dominant parameter since a surge of participants ($|\mathcal{I}|$) may crowd into a certain round, causing significant impacts overwhelming to any other parameters. The surge of participants might be closely dependent on external stimulates which are unknown to the system view.

We further investigate the concavity of the payoff function. Let $A=\frac{\partial^2 \mathbb{U}}{\partial \sigma^2}$, $B=\frac{\partial^2 \mathbb{U}}{\partial \sigma \partial r}$, $C=\frac{\partial^2 \mathbb{U}}{\partial r^2}$. Note that $d$ is an integer and always greater than 0, $\sigma$ and $r$ are both between 0 and 1, hence $AC-B^2 \ngeq 0$ and non-convex. This is an NP-hard problem that cannot be solved by typical convex optimization approaches. We omit the concrete exploration in this brief analysis but provide some possible solutions such as the Monte Carlo method or deep learning algorithms.

\section{Applications and Discussions} 
\label{sec_discussion}
\subsection{Applications and Opportunities}

\noindent\textbf{Fair Incentive Design.} The proposed rNFT scheme is particularly useful for scenarios where cross-references among NFTs are essential during the process of on-chain and off-chain profit sharing. A fair distribution mechanism must be established on top of existing \textit{observable} relationships including inheritance, reference, and extension. For example, an artist may develop his NFT work based on a previous NFT: a DJ may remix his record by referring to two pop songs, or a movie may include existing pieces of clips, etc. Explicit reference relationships for existing NFTs and enabling efficient queries on cross-references make much sense to the markets for renting, exchanging and trading.

\smallskip
\noindent\textbf{Broader Functionalities for NFT Markets.} NFT provides basic functionalities covering \textit{mint} and \textit{exchange}. This merely supports limited scenarios that are based on \textit{in-time} and \textit{one-time} trades. rNFT extends the sustainability and usability of tokens that enable consecutive transactions and long-term retrieving. positively fostering interaction between NFT creators, collectors, and enthusiasts.

\smallskip
\noindent\textbf{Promotion for More Integration.} Aligning with the rapid NFT development, our solution can follow and promote a series of emerging trends and capabilities. 

\noindent\hangindent 1em \textit{Mature technologies.} As discussed in Sec.\ref{sec-intro}, rNFT can be integrated with GNN for establishing NFT classification and predictions and with AI for plagiarism detection. Not surprisingly, rNFT can also be used to combine with traditional blockchain technologies such as off-chain (or layer-two \cite{gudgeon2020sok}) executions for portable usage, or cross-chain bridges for trading between different networks.

\noindent\hangindent 1em  \textit{Verifiable ownership\&provenance.}  rNFT can be used to securely store and share information about the origin and history of an asset, as well as providing strong connections for proof of ownership.

\noindent\hangindent 1em  \textit{DeFi/GameFi integration.} rNFT can be designed for more NFT and DeFi derivatives due to its strong attachments to previous history. The users/players will get trusted much more easily. Using rNFT to represent in-game items or to provide a new level of ownership and scarcity in gaming experiences is a promising direction.

\noindent\hangindent 1em  \textit{Physical goods and services.}  rNFT can help to Link NFTs to tangible goods and experiences, such as collectible merchandise, event tickets, and access to exclusive services, which is good to build a full view of user's profile.

\subsection{Challenges and Potential Solutions}
\noindent\textbf{Towards Cross-contract Invoke.} Our protocol is implemented based on a \textit{singly invoked} contract, which means that all minted NFTs should follow the same contract instructions. Each project can \textit{merely} establish a topological graph within its project, rather than arbitrarily other contracts that require cross-invoke. As in an initial stage, our preliminary implementation gives a demo for expanding the potential usage scope and facilitating the development of the following protocols/standards. To realize a cross-contract target, we plan to implement a series of additional functions that are backward compatible with several existing standards (e.g., EIP-998). A good start is that our single-invoke implementation can be seamlessly extended to the next cross-invoke stage. 

\smallskip
\noindent\textbf{Ways of Feeding Price.} The price oracle provides in-time prices feed to SC-enabled applications (e.g. DeFi protocols \cite{werner2022sok}). While many oracle designs have been implemented to provide data for fungible token prices, few attempts have been made to construct a price oracle  for NFTs. NFTs are generally less frequently traded, therefore creating difficulties for third parties to observe and retrieve price data. Our proposed scheme can normally operate without incentive indicators, independent of the price oracle. But with advanced incentive designs, price oracle is critical for users who make trades and obtain rewards. The price will impact both the initial price $p_0$ and round interest $r$, as well as will affect the behavioral trend from users. Embedding a secure and efficient oracle for feeding prices is of great importance.  

\smallskip
\noindent\textbf{Integrated Verification.} The rNFT-DAG in our protocol can \textit{merely} be used for reference only when rights need to be determined and certified without any compulsory measures in force in its current form. There could be risks that malicious rNFT publishers do not obey the protocol by not claiming the correct relationships with existing rNFT items, although the community could find out the mistakes in long-term observation.
A potential solution is to have an automated verification scheme integrated into our protocol where the peers have the ability to conduct mutual verification, as well as the incentive scheme associated with the malicious reference can be accordingly established.

\smallskip
\noindent\textbf{Deterministic Records with Low Latency.} The \textit{createdTimestamp} of each rNFT only covers the block-level timestamp (based on block headers), which does not support fine-grained comparisons such as transaction-level. The rNFT-DAG highly relies on the \textcolor{armygreen}{EIP-5521}-driven smart contracts that run on blockchain (Ethereum in our context). The block interval of a dozen seconds if the Proof-of-Work (PoW) is used by default needs to be reduced to achieve lower latency. This is significantly important when ones aim at preemptive registration of rNFTs. On the other hand, PoW is a probabilistic consensus algorithm that commonly causes a rollback of the state of smart contracts, which lengthens the waiting period until the reference relationship takes effect. A workaround is to use deterministic consensus algorithms such as the Proof-of-Authority (PoA) or any other Ethereum-compatible BFT algorithms (e.g., Quorum~\cite{Quorum_Blockchain_Service}).

\smallskip
\noindent\textbf{Complementary to CC0 License.} CC0 refers to ``Creative Commons 0'' -- a license that for the first time allows NFT holders to waive copyrights and other IP protections and significantly encourages the creation and extension of NFT derivatives associated with the original works~\cite{CC0}. This has been widely committed and adopted by many Web3 projects due to its attempt of eliminating the possible legal consequences associated with reference relationships between different NFT items and bring the items to a permissionless public domain -- the spirit of true Web3. However, it becomes difficult for one, especially those who are large-scale, well-established, or have a well-deserved reputation, to build commercial brands on top of a CC0-based NFT once opting for CC0 License as he/she has no right to exclude others from using the same origin (the CC0-based item). Our proposed rNFT protocol fills the gap for these NFT holders and can be considered complementary to the existing CC0 License due to its explicit indication of reference relationships.

\section{Conclusion}
\label{sec-conclusion}
In this paper, we propose a referable NFT (rNFT) scheme to improve the exposure and enhance the reference relationship of inclusive NFTs. We implement the scheme and propose the corresponding \textcolor{armygreen}{EIP-5521} standard to the community. We further establish the mathematical utility function to express the rNFT releasing and referring process. As far as we know, this is the first study that establishes a referable NFT network. By traversing it, a truly decentralized reference network comes to life to protect intellectual property and copyright and incentivizes creativity towards a fairer environment. 


\bibliographystyle{IEEEtran}
\bibliography{bib.bib}

\end{document}